\documentclass[aps,prl,twocolumn,showpacs,superscriptaddress]{revtex4-2}

\usepackage{graphicx}
\usepackage{amsmath}
\usepackage{multirow}
\usepackage{tabularx}
\usepackage{color}
\usepackage[colorlinks,linkcolor=blue,urlcolor=blue,anchorcolor=blue,citecolor=blue]{hyperref}

\begin{document}

\title{(Fe$_{1-x}$Ni$_{x}$)$_{5}$GeTe$_{2}$: an antiferromagnetic triangular Ising lattice with itinerant magnetism}
	
\author{Xunwu Hu}
\author{Dao-Xin Yao}
\email{yaodaox@mail.sysu.edu.cn}
\author{Kun Cao}
\email{caok7@mail.sysu.edu.cn}
\affiliation{Center for Neutron Science and Technology, Guangdong Provincial Key Laboratory of Magnetoelectric Physics and Devices, State Key Laboratory of Optoelectronic Materials and Technologies, School of Physics, Sun Yat-Sen University, Guangzhou, 510275, China}

\begin{abstract}
Based on first-principles calculations, an antiferromagnetic Ising model on a triangular lattice has been proposed to interpret the order of Fe$(1)$-Ge pairs and the formation of $\sqrt{3}$ $\times $ $\sqrt{3}$ superstructures in the Fe$_5$GeTe$_2$ (F5GT), as well as to predict the existence of similar superstructures in Ni doped F5GT (Ni-F5GT). Our study suggests that F5GT systems may be considered as a structural realization of the well known antiferromagnetic Ising model on a triangular lattice. Based on the superstructures, a Heisenberg-Landau Hamiltonian, taking into account both Heisenberg interactions and longitudinal spin fluctuations, is implemented to describe magnetism in both F5GT and Ni-F5GT. We unveil that frustrated magnetic interactions associated with Fe(1), tuned by a tiny Ni doping ($x \leq 0.2 $), is responsible for the experimentally observed enhancement of the $T_c$ to 478 K in Ni-F5GT. Our calculations show that at low doping levels, monolayer Ni-F5GT has almost the same magnetic phase diagram as that of the bulk, which indicates a pervasive beyond room temperature ferromagnetism in this Ni doped two-dimensional system.
\end{abstract}
\maketitle
{\it Introduction.}$-$In recent years, quasi-two-dimensional (quasi-2D) van der waals (vdW) ferromagnetic (FM) materials, such as Cr$_{2}$Ge$_{2}$Te$_{6}$\cite{2017gong}, CrI$_{3}$\cite{2017huan}, CrTe$_{2}$\cite{2015jpcm,2020sun} and Fe$_{n}$GeTe$_{2}$ (${n}$ = 3, 4, 5)\cite{2018deng,2020sciadv,2019acsnano,2018za}, have garnered significant attention due to their novel physical properties and potential applications in spintronic devices. Among them, Fe$_5$GeTe$_2$ (F5GT) system is particularly attractive, owing to its extraordinary properties suitable for practical applications, including metallic ferromagnetism with near room temperature $T_c$ (279-332 K), large saturation moment (1.8-2.1 $\mu_B$/Fe) \cite{2018za,2019acsnano,2020prbzhang,20212dm,chen2022} and high tunability of chemical compositions\cite{2018za,2021may,2022prl,2022sciav,2022prmz}. More recently, Chen {\it et al.} reported that, by 36 $\%$ Ni doping in the F5GT compound, its $T_c$ can be significantly enhanced to 478 K\cite{2022prl}. They further postulated that this enhancement of $T_c$ in Ni doped F5GT (Ni-F5GT) may be caused by increased magnetic exchange interactions due to the structural alterations.

On the other hand, one of the most important characteristics of the F5GT systems is that there are two split sites of Fe positions, labelled here as Fe(1)$_{up}$ and Fe(1)$_{dn}$, with only one position occupiable within each Fe(1)$_{up}$, Fe(1)$_{dn}$ pair, leading to intrinsic structural disorders\cite{2019acsnano}. Due to the difficulty of dealing with disorder, so far, most theoretical studies are based on an ordered lattice structure with fully occupied Fe(1)$_{up}$($uuu$) or Fe(1)$_{dn}$ ($ddd$) positions\cite{2019first,2021zhaoyu,2022prmz,2022liu,2021prbyao}. Intriguingly, scanning tunneling microscopy (STM) experiments indicate that the F5GT compound has ordered $\sqrt{3}$ $\times$ $\sqrt{3}$ superstructures driven by the ordering of Fe$(1)$ atoms\cite{2021adfm,2021prb}. It is then natural to suspect that lattice structural details may have subtle interplay with electronic structure and magnetic ordering in both F5GT and Ni-F5GT and the $\sqrt{3}$ $\times$ $\sqrt{3}$ superstructures could be more representable to the true experimental lattice structures. Ershadrad {\it et al.} investigated the structural ordering of F5GT based on ${ab}$ ${initio}$ calculated energies\cite{2022jpcl}, but ignored the influence of temperature and entropy. Therefore, the mechanism behind the formation of the $\sqrt{3}$ $\times$ $\sqrt{3}$ superstructures and their interplay with magnetic ordering, especially the high $T_c$'s, remain elusive.

\begin{figure}[t]
	\includegraphics[scale=0.35]{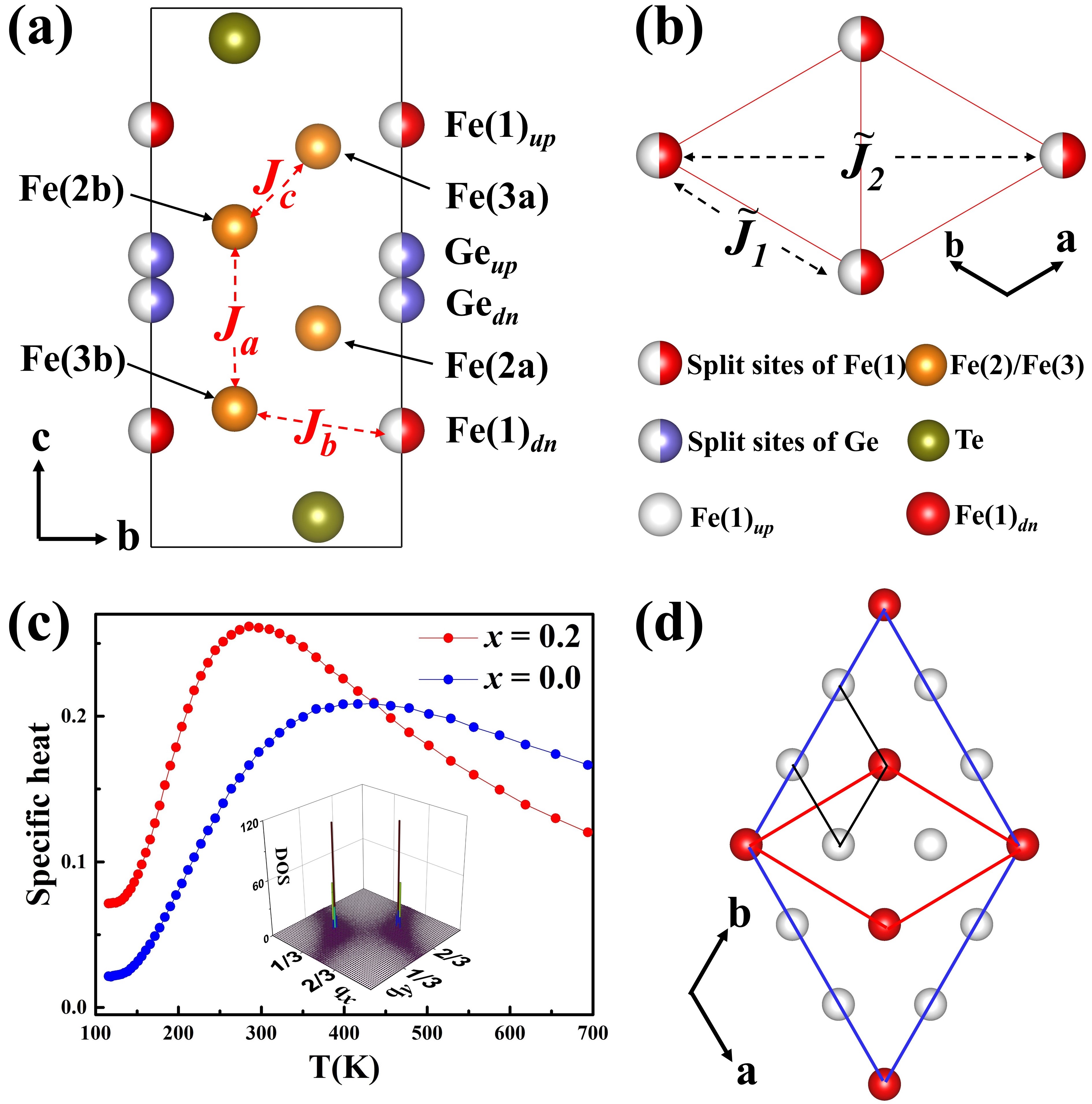}
	\caption{(a) Average crystal structure of F5GT sublayer. The white, red, orange, blue and dark green spheres represent Fe(1)$_{up}$, Fe(1)$_{dn}$, Fe(2)(3), Ge, and Te atoms, respectively. (b) Illustration of Fe(1) sublattice, half white and red indicate split sites of Fe(1) on a triangular lattice. (c) Specific heat as a function of temperature simulated for $x=0$  and $x=0.2$. The insert depicts the analysis of the discrete Fourier transformation of the structure at 153 K for $x=0$. (d) Schematic of the SRO superstructure. The $3\times3$ superstructure is marked by the blue axis, with the corresponding primitive $\sqrt{3}\times\sqrt{3}$ superstructure marked by the red axis.}
	\label{fig1}
\end{figure}

In this letter, we propose an antiferromagnetic Ising model on a triangular lattice, based on first principles calculations, to interpret the ordering of Fe$(1)$ atoms and the formation of the $\sqrt{3}$ $\times $ $\sqrt{3}$ superstructures in F5GT. Similar superstructures are then predicted to exist in Ni-F5GT. Our study suggests that F5GT systems may be considered as a structural realization of the well known antiferromagnetic Ising model on a triangular lattice. Based on the superstructures, a Heisenberg-Landau Hamiltonian, taking into account longitudinal spin fluctuations, is implemented to study the magnetic properties of F5GT and Ni-F5GT. We unveil that frustrated magnetic interactions associated with Fe(1), tuned by a tiny Ni doping, is responsible for the experimentally observed significant enhancement of the $T_c$ in Ni-F5GT. In contrast, calculated results based on the $uuu$ structure deviate qualitatively from experiments. Our calculations further show that at low doping levels, the magnetic behaviors of the monolayer Ni-F5GT resemble closely that of the bulk, therefore indicates pervasive beyond-room-temperature two-dimensional ferromagnetism.


{\it Lattice structure.}$-$The average crystal structure of bulk F5GT has a rhombohedral space group R$\bar{3}$m \cite{2019acsnano}, with three identical layers stacked staggeringly in each unit cell, labeled as ABC stacking. Each layer consists of a Fe$_{5}$Ge sublayer sandwiched between two Te planes, as shown in Fig. \ref{fig1}(a). The two split-sites of Fe$(1)$ atoms are in the outermost plane of each Fe$_{5}$Ge sublayer. First principles calculations indicate that Fe(1)$_{up}$-Ge$_{up}$ or Fe(1)$_{dn}$-Ge$_{dn}$ pairs are not allowed due to bond distance restrictions, i.e. when Fe$(1)$ atom occupies the up site, the corresponding Ge atom can only occupy the down site, forming Fe(1)$_{up}$-Ge$_{dn}$ or equivalently Fe(1)$_{dn}$-Ge$_{up}$ pairs. Interestingly, Fe$(1)$ positions form a triangular lattice in the plane, as shown in Fig. \ref{fig1}(b). If Fe(1)$_{up}$ and Fe(1)$_{dn}$ are viewed as spin up and spin down, respectively, the problem of Fe(1) order translates into an Ising model on a triangular lattice. The Hamiltonian of the Ising model can then be written as
\begin{equation}
H=\sum\limits_{i<j}\tilde{J}_{ij}S_{i}\cdot S_{j}\quad(S_{i,j}=\pm1)
	\label{eq1}
\end{equation}
where $\tilde{J}_{i}$ corresponds to the $i$th nearest neighbour (NN) interaction(see Fig.~\ref{fig1}(b)). We calculate $\tilde{J}_{i}$ up to the second nearest neighbour by fitting to first principles calculated energies, obtaining $\tilde{J}_{1}$ = 30.2 meV and $\tilde{J}_{2}$ = 0.99 meV for F5GT, which shows that this system is an antiferromagnetic frustrated Ising model on a triangular lattice.  Such a model with only $\tilde{J}_{1}$ is well known for its lack of long range order at finite temperatures with an infinitely degenerate ground state, comprised of $\sqrt{3}$ $\times$ $\sqrt{3}$ superstructures with two up and one down spins (or equivalently two down and one up) on each triangle{\cite{1950prv,1984jpsj}}. The additional small $\tilde{J}_{2}$ may break the infinite degeneracy, but barely change the general behavior of the original model at the temperatures that we are concerned (see our simulations in Fig. S2). Specific heat from re-MC simulations show a bump peaked at around 380 K, which does not become sharper with increasing lattice size (shown in Fig. S2(a)), indicating a weak size dependence and short range ordered (SRO) structures (see Fig. \ref{fig1}(c) blue curve) \cite{1984jpsj}. Spectra analysis by performing discrete Fourier transformation on the structures obtained from re-MC at low temperatures shows the highest density of states locating at $q=(1/3,1/3, 0)$ and $q=(2/3,2/3, 0)$(inset of Fig. \ref{fig1}(c)), indicating that the SRO structures are, indeed, $\sqrt{3}$ $\times$ $\sqrt{3}$ superstructures, which is in good agreement with the fast Fourier transformation pattern of the high-angle annular dark-field (HAADF) images from experiments\cite{2019acsnano}. A representative superstructure is shown in Fig.~\ref{fig1}(d). Notably, the superstructure preserve the C$_{3}$ rotational symmetry of the averaged structure, exhibiting a Fe(1)$_{dn}$Fe(1)$_{up}$Fe(1)$_{up}$ ($duu$) order along all three [100], [010] and [110] directions.

\begin{figure}[t]
	\includegraphics[scale=0.36]{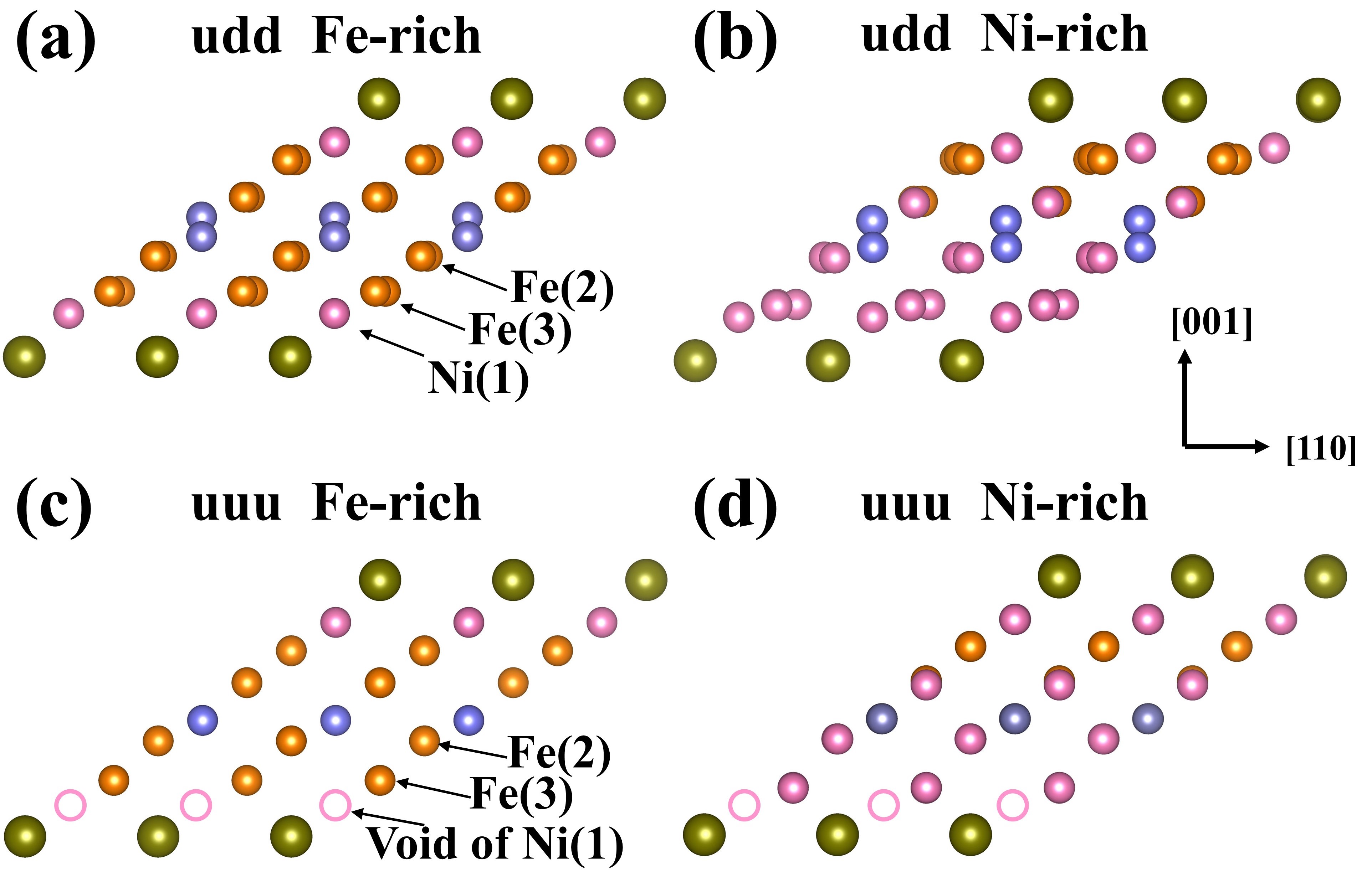}
    \centering
	\caption{Illustrations of the TGT plane structure in Ni-F5GT systems. The pink, orange, blue and dark green spheres represent Ni, Fe, Ge, and Te atoms, respectively. The optimized duu structure at (a)$x = 0.2$. (b)$x = 0.667$. The optimized uuu structure at (c)$x = 0.2$. (d)$x = 0.667$.}
	\label{fig2}
\end{figure}

\begin{figure}[t]
	\includegraphics[scale=0.31]{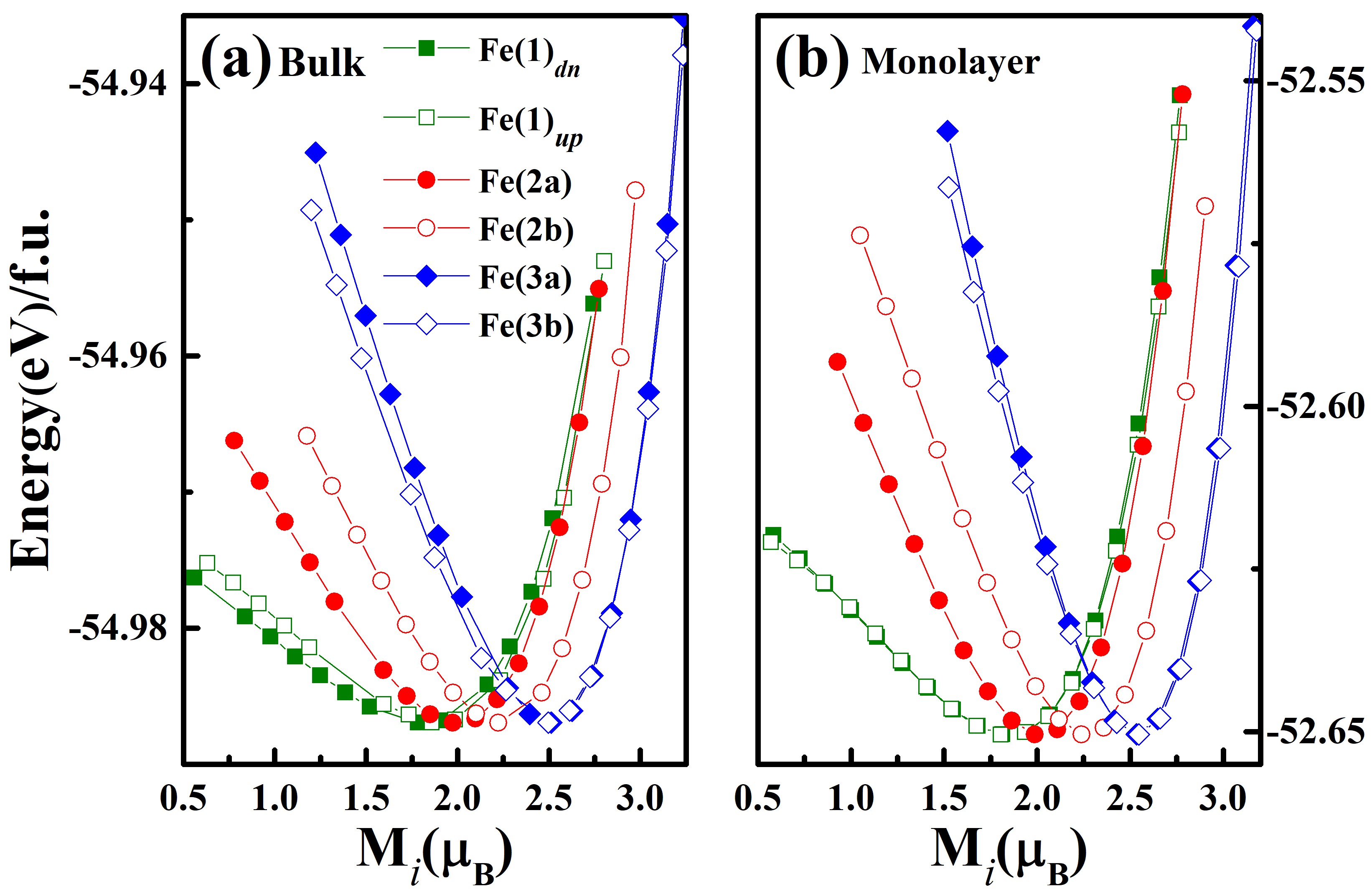}
    \centering
	\caption{${\bf M}_{i}$-dependent energy in bulk(a) and monolayer(b) F5GT.}
	\label{fig3}
\end{figure}

For Ni-F5GT, although the average structure with Ni doping changes from ABC stacking to AA stacking (space group P$\bar{3}$m1)\cite{2022prl}, the geometry of each sublayer still keeps intact. Therefore, the intrinsic structural disorder induced by split sites of Fe(1) occupation persists with Ni doping, with a possibility of forming similar superstructures as in F5GT. To make our computations feasible, the Ni doping is simulated by using the lowest energy structure obtained at each corresponding Ni content in a supercell. Our calculations show that the doped Ni atoms tend to first substitute the Fe(1), then gradually replace Fe(2) and Fe(3) with increasing Ni content.
We here use Ni-F5GT with $x = 0.2$, in which case all Fe(1) atoms are replaced by Ni atom with other Fe atoms unaffected, as an example to illustrate the potential structural ordering in Ni-F5GT. The same Ising model (Eq.~\ref{eq1}) can then be applied, with $\tilde{J}_{ij}$ now standing for the interaction between Ni-Ge structural pairs. We obtain $\tilde{J}_{1}$ = 17.9 meV, $\tilde{J}_{2}$ = -3.2 meV, with simulated specific heat shown in Fig. \ref{fig1}(c) (red curve). Compared to $x = 0$, the bump of the specific heat becomes sharper, with its peak position shifted to a lower temperature at around 300 K, as a result of smaller $\tilde{J}$ values, but larger $\tilde{J}_{2}/ \tilde{J}_{1}$ ratio, making it deviating more from a perfect triangular NN antiferromagnetic Ising model. The results of corresponding Fourier spectra analysis of Ni-F5GT with $x = 0.2$ is similar to that of F5GT, showing two peaks at $q=(1/3,1/3, 0)$ and $q=(2/3,2/3, 0)$, but with higher peak intensity, which indicates that with Ni doping, the $\sqrt{3}$ $\times$ $\sqrt{3}$ superstructure should also be observable (see Fig. S3). To examine more of the structure, in each sublayer of F5GT, the sites of Fe atoms are symmetric with respect to the Ge atom at the center, forming a Te-Fe(1)-Fe(3)-Fe(2)-Ge-Fe(2)-Fe(3)-Fe(1)-Te plane (labeled as TGT plane) (See Fig. \ref{fig2}(a)). First principles optimized $\sqrt{3}\times \sqrt{3}$ superstructure of Ni-F5GT (duu) with $x = 0.2$ shows flat TGT plane (see Fig. \ref{fig2}(a)), while for $x = 0.667$, the TGT plane is rumpled, which both agree perfectly with HAADF scanning transmission electron microscopy experiments\cite{2022prl}. In contrast, structure optimised with the Fe(1)$_{up}$Fe(1)$_{up}$Fe(1)$_{up}$ (uuu) structure of Ni-F5GT can not accommodate a complete TGT plane, with the lost of one Fe(1)/Ni atom, due to that all Fe(1) atoms occupy the up site, further supporting the existence of the $\sqrt{3}$ $\times$ $\sqrt{3}$ superstructure in Ni-F5GT (see Figs. \ref{fig2}(c)(d)). It is therefore reasonable to argue that at finite temperatures, the experimental lattice structures are comprised of domains with different SRO $\sqrt{3}$ $\times$ $\sqrt{3}$ superstructures.
\begin{figure}[t]
	\includegraphics[scale=0.31]{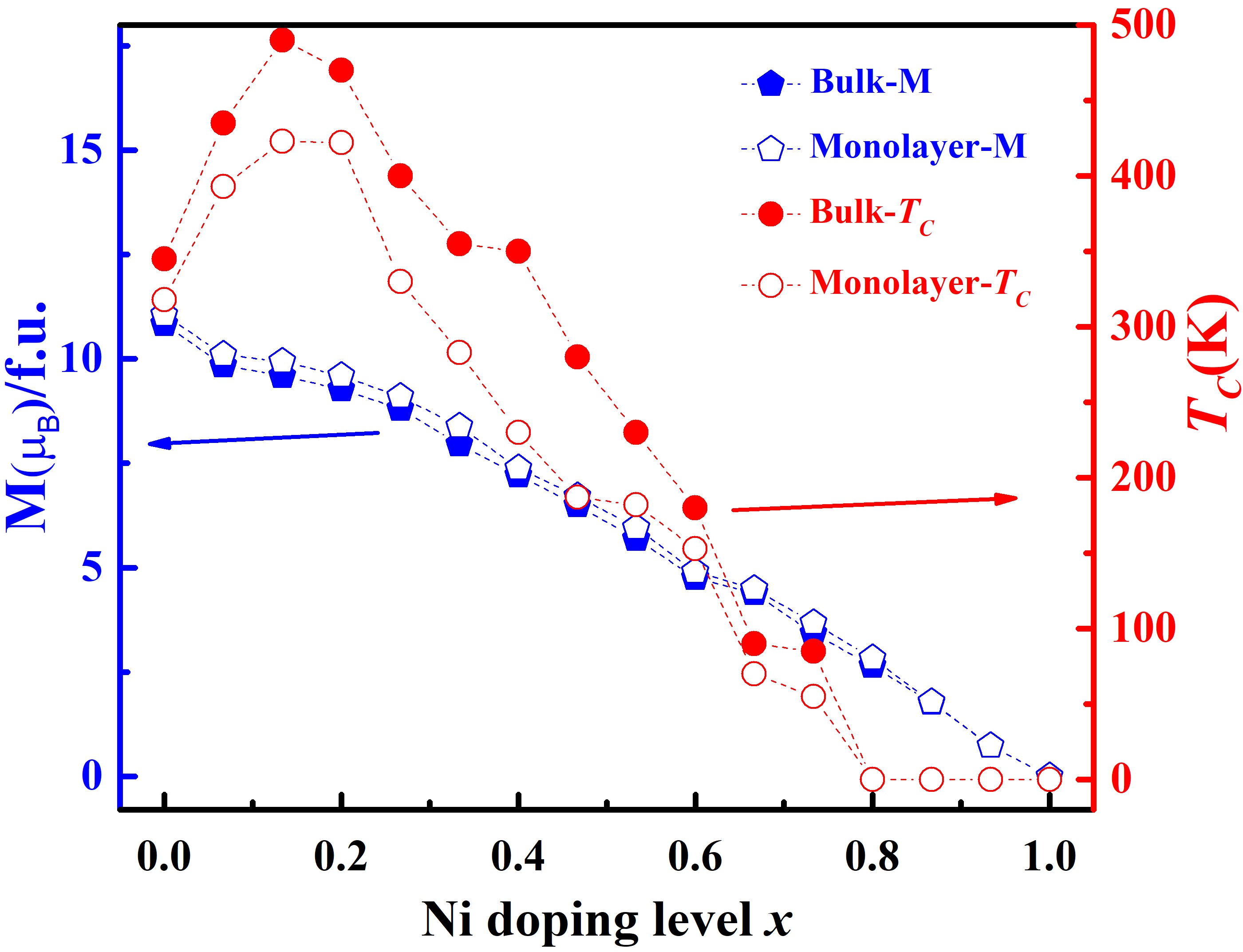}
   \centering
	\caption{Ni doping level $x$-dependent magnetic moments and $T_c$ of bulk (solid symbol) and monolayer (open symbol) Ni-F5GT systems marked by blue and red curves, respectively.}
	\label{fig4}
\end{figure}

{\it Magnetism.}$-$It is straightforward to assume that magnetism of the ordered superstructure could exhibit noticeable difference from that of the uuu (or ddd) structure, due to the change of local environment related to Fe(1) atoms, as also suggested in Ref.~\cite{2022jpcl}. We therefore further investigate the magnetism, firstly the magnetic moments ${\bf M}$, using the ordered superstructure. For F5GT, Fe atoms are divided into six inequivalent positions, labelled here as Fe(1)$_{up}$, Fe(1)$_{dn}$, Fe(2a), Fe(2b), Fe(3a), Fe(3b) {(see Fig. \ref{fig1}(a))}, with each $|\textbf M|$ of the bulk calculated as 1.80, 1.88, 1.98, 2.24, 2.49 and 2.45 $\mu_B$, respectively, which agree with the values 0.6 - 2.6 $\mu_B$/Fe obtained from experiments\cite{2019acsnano}. With Ni doping, the calculated total magnetic moments $|{\bf M}_{tot}|$ of the bulk Ni-F5GT fall approximately linearly with the increase of doping level $x$, from $10.8 \mu_{B}/f.u.$ at $x = 0$ to zero at $x = 1$ (see Fig. \ref{fig4}). The doped Ni atoms are found to be weakly magnetic with ${\bf M}_{Ni}\sim 0.27 \mu_B$ at $x \leq 0.5$, , while at higher doping levels, ${\bf M}_{Ni}$ vanishes, therefore result in a progressive dilution of the total magnetization with the increasing of $x$. For more details, the magnetic moment at each different site decrease stagewise because of the nearly ordered substitution of each Fe site by Ni, i.e. fully replace one site before going to the next (see Fig. S6). When it comes to the monolayer, the $|{\bf M}_{tot}|$ behaves similarly with that of the bulk, but have slightly larger magnitude because of enhanced localization in the low-dimensional case.

Furthermore for F5GT, we study the change of each $|\bf M|$ by varying their relative canting angles. For instance, when spin angle is forced from $0^{\circ }$ (parallel to other spins) to $90^{\circ }$ (perpendicular to other spins), $|${\bf M}$_1|$, $|${\bf M}$_{2a}|$ and $|${\bf M}$_{2b}|$ are each reduced by a non-negligible $0.6$ $\mu_B$, while $|${\bf M}$_{3a}|$ and $|${\bf M}$_{3b}|$ remain nearly constant (see Fig. S4.), which indicate the coexistence of itinerant and localized magnetism in the F5GT systems, with the itinerant character mainly situated on Fe(1), Fe(2a) and Fe(2b) atoms. Moreover, we calculate the total energy dependence on the variation of each $|${\bf M}$_i|$, with results shown in  Fig. \ref{fig3}. It can be seen that the energy wells on $|${\bf M}$_1|$, $|${\bf M}$_{2a}|$ and $|${\bf M}$_{2b}|$ are significantly asymmetric with respect to each well bottom, with a flatter rising on the low moment side, typical for itinerant magnetic systems, such as bcc Fe and fcc Ni{\cite{prbma,2007prbr,97prbr,97prl}}. This could result in noticeable redistribution of relevant $|${\bf M}$_i|$ with variation of temperature (See Fig. S10), even lead to finite magnetic moments purely stabilized by thermal fluctuations at high temperatures.  However, the energy wells on the $|${\bf M}$_{3a}|$ and $|${\bf M}$_{3b}|$ are much deeper and nearly symmetric with respect to each well bottom, therefore $|${\bf M}$_{3a}|$ and $|${\bf M}$_{3b}|$ are expected to remain localized and only exhibit negligible fluctuations in the temperature range we are interested in. When Ni doping is introduced, each remaining Fe magnetic moment largely keeps their original character. However, with only small magnetic moments around  0.27$\mu_B$, {\bf M}$_{Ni}$ is also expect to be strongly itinerant (see Figs. S5 and S6).

To describe simultaneously the localized and itinerant magnetism in the F5GT systems, we resort to the Heisenberg-Landau Hamiltonian {\cite{prbma,2007prbr,97prbr,97prl}},
\begin{eqnarray}
H &=&\sum\limits_{i<j}J_{ij}{\bf M}_{i}\cdot {\bf M}_{j}-\sum\limits_{i}D\left( {\bf M}_{i}^{z}\right)^{2}\notag \\
 && +\sum\limits_{i}A_{i}{\bf M}_{i}^{2}+\sum\limits_{i}B_{i}{\bf M}_{i}^{4}+\sum\limits_{i}C_{i}{\bf M}_{i}^{6}
\end{eqnarray}
where the first two terms represent Heisenberg exchange interactions with single ion anisotropy and the rest correspond to the Landau expansion terms, taking into account the longitudinal fluctuation of magnetic moments. For simplicity, we choose $J_{ij}$ to be $\bf M$ independent and the Landau terms are expanded to the sixth order. The expansion coefficients $A,B$ and $C$ are only considered for magnetic moments with noticeable itinerant characters, $|${\bf M}$_1|$, $|${\bf M}$_{2a}|$, $|${\bf M}$_{2b}|$ and $|${\bf M}$_{Ni}|$. For ${\bf M}_{Ni}$, an 8$th$ term $\sum_{i}D_{i}{\bf M}_{Ni}^{8}$ is also added for a better fit. The $J's$ are then calculated using the TB2J code\cite{2021cpc} and the expansion coefficients are calculated by fitting to first principles constrained magnetic moment calculations, with obtained results shown in Tables S2-S5. Re-MC simulations are then performed to study phase diagram of this Hamiltonian.  Only single phase transitions from paramagnetic (PM) phases to FM phases are obtained , with $T_c$ of bulk and monolayer F5GT at about 345 K and 318 K, respectively, which are slightly higher than experimental values 279-332 K for bulk and 280-300 K for monolayer\cite{2019acsnano,2020prbzhang,20212dm,chen2022}. It is worth noting that experimental samples of F5GT all have noticeable Fe deficiencies, which could lead to lower $T_c$ compared to simulations on perfect crystals. Surprisingly, the longitudinal spin fluctuations are found to only affect the $T_c$'s mildly, with a modification of $\sim 5 \%$ with respect to that obtained with standard Heisenberg interactions (See Fig. S9), less significant than the $\sim 10 \%$ lowering revealed for bcc Fe \cite{prbma}. This rather weak effect from the Landau terms is due to that only part of the Fe magnetism can be considered itinerant and the dominant FM exchange interactions are not heavily affected. Notably, our calculations on bulk Ni-F5GT also show that a light Ni doping can enhance the $T_c$, up to 490 K at $x = 0.2$ and then slowly reduce the $T_c$, down to zero when $x > 0.8$, exhibiting overall behavior in good agreement with recent experimental observations\cite{2022prl}.  Moreover, our simulations show that at low doping levels, $T_c$ of monolayer Ni-FGT is only slightly lower that its counterpart of the bulk, indicating pervasive ferromagnetism in 2D Ni-F5GT.

\begin{figure}[t]
	\includegraphics[scale=0.34]{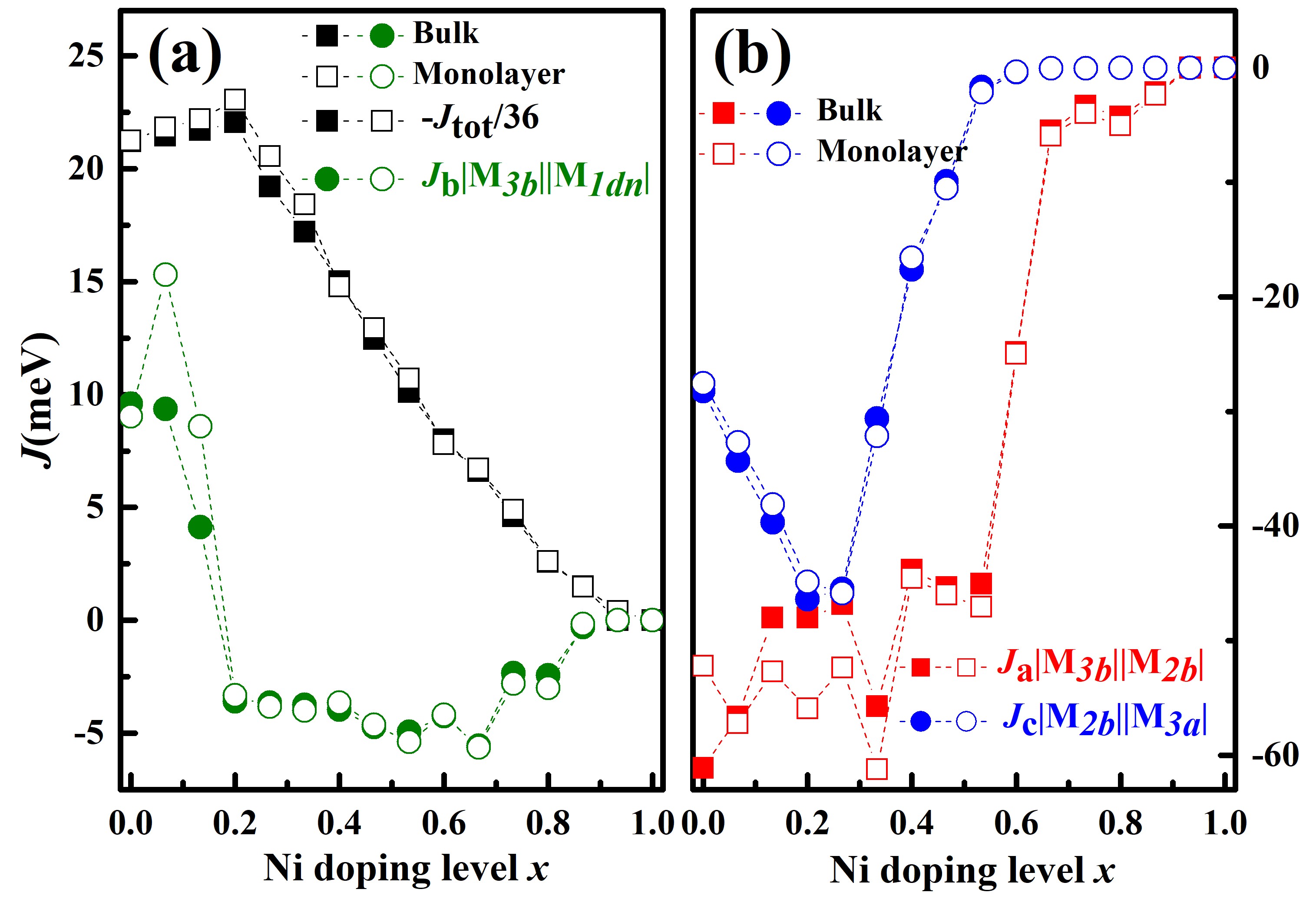}
	\caption{The major exchange interactions of bulk (solid symbol) and monolayer (open symbol) of Ni-F5GT as a function of Ni doping level $x$.}
	\label{fig5}
\end{figure}

We now look into the calculated $J_{ij}$ to explore the underlying mechanism behind the tuning of $T_c$ by Ni doping. As a first step, we look at the sum of all $J's$ with $J_{tot}=\frac{1}{2}\sum_{ij}J_{ij}$$|${\bf M}$_{i}$$|$$|${\bf M}$_{j}|$, which is roughly proportional to $T_c$ on the mean field level.
Not out of expectation, it can be seen that $J_{tot}$ is strongly negative with its absolute values following the same trend with Ni doping as that of the calculated $T_c$ (Fig. \ref{fig5}(a)). When moderate Ni doping is introduced($0$ $\leq x\leq 0.53$), the strongest intralayer interaction $J_{a}$$|${\bf M}$_{2b}|$$|${\bf M}$_{3b}|$ (marked in Fig. \ref{fig1}(a)) remains ferromagnetic ($-45$ $\sim -60$ meV), providing dominant contribution to stabilize the FM phase. As aforementioned, under light Ni doping with $x \leq 0.2$, only Fe(1) is substituted, therefore in this case, exchange interactions related to Fe(1) affect the tuning of $T_c$ directly. In the case of zero doping, the exchange interactions related to Fe(1) are frustrated, with AFM interactions between Fe(1) and its nearest Fe(3b)($J_{b}$$|${\bf M}$_{3b}|$$|${\bf M}$_{1dn}|$ =  9.6 meV), but stronger FM interactions between Fe(1) and Fe(2a) ($J_{1,2a}$$|${\bf M}$_{1dn}|$$|${\bf M}$_{2a}|$
= -24.66 meV), therefore Fe(1) is forced to be parallel to Fe(2a) to stabilize the FM state. A light Ni doping replaces the original magnetic moment of Fe(1)($\sim$1.8 $\mu_B$) by smaller $M_{Ni}$($\sim$ 0.27 $\mu_B$), resulting in a decrease of average AFM $J_b$, which even become FM with $x$ $\geq$ 0.2. On the other hand, as shown in Fig. \ref{fig5}(b), the overall strength of FM $J_c$$|${\bf M}$_{2b}|$$|${\bf M}$_{3a}|$ is greatly promoted from -28 to -48 meV at $0 \leq x \leq 0.2$. This cooperative weakening of AFM $J_b$ and the enhancement of FM $J_c$ thus play the pivotal role to increase the $T_c$ at low doping level. However, further Ni doping can dilute major FM couplings (see Fig. \ref{fig5}b), leading to the decrease of $T_c$.

For comparison, we also investigate magnetism of F5GT and Ni-F5GT using the $uuu$ structure, with calculated magnetic moments and $T_c$ shown in Fig. S11. In this case, the magnetic moments of Fe(1) vanish. Therefore, with light Ni doping, the saturated magnetic moments are calculated to rise first and then decrease, in qualitative variation with the monotonous decreasing behavior of corresponding experimental results. Although Curie temperatures are calculated to be similarly enhanced under low level doping, but is up to 525 K at $x=0.2$, which is much higher than the experimental values. It is worth emphasizing that our focus in this work, is on the itinerant magnetism and $T_c$, other than the seemingly intricate phase diagrams observed below $T_c$ in experiments, which may well be related to weak, but subtle interlayer interactions, since a tiny Ni doping induced structural transition from ABC stacking to AA stacking can completely smear the complexity out and bring the phase diagram to a simple single PM to FM transition (see Ref.\cite{2022prl}). To capture these subtle interactions may be beyond the current capabilities of standard {\it ab initio} calculations and beyond the scope of this work, we thus leave it to future studies.

In summary, we propose an antiferromagnetic Ising model on a triangular lattice, based on first principles calculations, to interpret the ordering of Fe$(1)$ atoms and the formation of the $\sqrt{3}$ $\times $ $\sqrt{3}$ superstructures in F5GT. Similar superstructures are then predicted to exist in Ni-F5GT. Our study suggests that F5GT systems may be considered as a structural realization of the well known antiferromagnetic Ising model on a triangular lattice. Based on the superstructures, a Heisenberg-Landau Hamiltonian, taking into account longitudinal spin fluctuations, is implemented to study the magnetic properties of F5GT and Ni-F5GT. We unveil that frustrated magnetic interactions associated with Fe(1), tuned by a tiny Ni doping, is responsible for the experimentally observed significant enhancement of the $T_c$ in Ni-F5GT. In contrast, calculated results based on the $uuu$ structure deviate qualitatively from experiments. Our calculations further show that at low doping levels, the magnetic behaviors of the monolayer Ni-F5GT resemble closely that of the bulk, therefore indicates pervasive beyond-room-temperature two-dimensional ferromagnetism.

Work at Sun Yat-Sen University was supported by the National Key Research and Development Program of China (Grants No. 2018YFA0306001, 2017YFA0206203), and the Guangdong Basic and Applied Basic Research Foundation (Grants No. 2022A1515011618, No. 2019A1515011337), and the National Natural Science Foundation of China (Grants No. 92165204, No. 11974432), Shenzhen International Quantum Academy (Grant No. SIQA202102), Leading Talent Program of Guangdong Special Projects (201626003).

\bibliography{reference}

\begin{thebibliography}{29}%
\makeatletter
\providecommand \@ifxundefined [1]{%
 \@ifx{#1\undefined}
}%
\providecommand \@ifnum [1]{%
 \ifnum #1\expandafter \@firstoftwo
 \else \expandafter \@secondoftwo
 \fi
}%
\providecommand \@ifx [1]{%
 \ifx #1\expandafter \@firstoftwo
 \else \expandafter \@secondoftwo
 \fi
}%
\providecommand \natexlab [1]{#1}%
\providecommand \enquote  [1]{``#1''}%
\providecommand \bibnamefont  [1]{#1}%
\providecommand \bibfnamefont [1]{#1}%
\providecommand \citenamefont [1]{#1}%
\providecommand \href@noop [0]{\@secondoftwo}%
\providecommand \href [0]{\begingroup \@sanitize@url \@href}%
\providecommand \@href[1]{\@@startlink{#1}\@@href}%
\providecommand \@@href[1]{\endgroup#1\@@endlink}%
\providecommand \@sanitize@url [0]{\catcode `\\12\catcode `\$12\catcode
  `\&12\catcode `\#12\catcode `\^12\catcode `\_12\catcode `\%12\relax}%
\providecommand \@@startlink[1]{}%
\providecommand \@@endlink[0]{}%
\providecommand \url  [0]{\begingroup\@sanitize@url \@url }%
\providecommand \@url [1]{\endgroup\@href {#1}{\urlprefix }}%
\providecommand \urlprefix  [0]{URL }%
\providecommand \Eprint [0]{\href }%
\providecommand \doibase [0]{https://doi.org/}%
\providecommand \selectlanguage [0]{\@gobble}%
\providecommand \bibinfo  [0]{\@secondoftwo}%
\providecommand \bibfield  [0]{\@secondoftwo}%
\providecommand \translation [1]{[#1]}%
\providecommand \BibitemOpen [0]{}%
\providecommand \bibitemStop [0]{}%
\providecommand \bibitemNoStop [0]{.\EOS\space}%
\providecommand \EOS [0]{\spacefactor3000\relax}%
\providecommand \BibitemShut  [1]{\csname bibitem#1\endcsname}%
\let\auto@bib@innerbib\@empty
\bibitem [{\citenamefont {Gong}\ \emph {et~al.}(2017)\citenamefont {Gong},
  \citenamefont {Li}, \citenamefont {Li}, \citenamefont {Ji}, \citenamefont
  {Stern}, \citenamefont {Xia}, \citenamefont {Cao}, \citenamefont {Bao},
  \citenamefont {Wang}, \citenamefont {Wang}, \citenamefont {Qiu},
  \citenamefont {Cava}, \citenamefont {Louie}, \citenamefont {Xia},\ and\
  \citenamefont {Zhang}}]{2017gong}%
  \BibitemOpen
  \bibfield  {author} {\bibinfo {author} {\bibfnamefont {C.}~\bibnamefont
  {Gong}}, \bibinfo {author} {\bibfnamefont {L.}~\bibnamefont {Li}}, \bibinfo
  {author} {\bibfnamefont {Z.}~\bibnamefont {Li}}, \bibinfo {author}
  {\bibfnamefont {H.}~\bibnamefont {Ji}}, \bibinfo {author} {\bibfnamefont
  {A.}~\bibnamefont {Stern}}, \bibinfo {author} {\bibfnamefont
  {Y.}~\bibnamefont {Xia}}, \bibinfo {author} {\bibfnamefont {T.}~\bibnamefont
  {Cao}}, \bibinfo {author} {\bibfnamefont {W.}~\bibnamefont {Bao}}, \bibinfo
  {author} {\bibfnamefont {C.}~\bibnamefont {Wang}}, \bibinfo {author}
  {\bibfnamefont {Y.}~\bibnamefont {Wang}}, \bibinfo {author} {\bibfnamefont
  {Z.~Q.}\ \bibnamefont {Qiu}}, \bibinfo {author} {\bibfnamefont {R.~J.}\
  \bibnamefont {Cava}}, \bibinfo {author} {\bibfnamefont {S.~G.}\ \bibnamefont
  {Louie}}, \bibinfo {author} {\bibfnamefont {J.}~\bibnamefont {Xia}},\ and\
  \bibinfo {author} {\bibfnamefont {X.}~\bibnamefont {Zhang}},\ }\bibfield
  {title} {\bibinfo {title} {Discovery of intrinsic ferromagnetism in
  two-dimensional van der waals crystals},\ }\href
  {https://doi.org/10.1038/nature22060} {\bibfield  {journal} {\bibinfo
  {journal} {Nature}\ }\textbf {\bibinfo {volume} {546}},\ \bibinfo {pages}
  {265} (\bibinfo {year} {2017})}\BibitemShut {NoStop}%
\bibitem [{\citenamefont {Huang}\ \emph {et~al.}(2017)\citenamefont {Huang},
  \citenamefont {Clark}, \citenamefont {Navarro-Moratalla}, \citenamefont
  {Klein}, \citenamefont {Cheng}, \citenamefont {Seyler}, \citenamefont
  {Zhong}, \citenamefont {Schmidgall}, \citenamefont {McGuire}, \citenamefont
  {Cobden}, \citenamefont {Yao}, \citenamefont {Xiao}, \citenamefont
  {Jarillo-Herrero},\ and\ \citenamefont {Xu}}]{2017huan}%
  \BibitemOpen
  \bibfield  {author} {\bibinfo {author} {\bibfnamefont {B.}~\bibnamefont
  {Huang}}, \bibinfo {author} {\bibfnamefont {G.}~\bibnamefont {Clark}},
  \bibinfo {author} {\bibfnamefont {E.}~\bibnamefont {Navarro-Moratalla}},
  \bibinfo {author} {\bibfnamefont {D.~R.}\ \bibnamefont {Klein}}, \bibinfo
  {author} {\bibfnamefont {R.}~\bibnamefont {Cheng}}, \bibinfo {author}
  {\bibfnamefont {K.~L.}\ \bibnamefont {Seyler}}, \bibinfo {author}
  {\bibfnamefont {D.}~\bibnamefont {Zhong}}, \bibinfo {author} {\bibfnamefont
  {E.}~\bibnamefont {Schmidgall}}, \bibinfo {author} {\bibfnamefont {M.~A.}\
  \bibnamefont {McGuire}}, \bibinfo {author} {\bibfnamefont {D.~H.}\
  \bibnamefont {Cobden}}, \bibinfo {author} {\bibfnamefont {W.}~\bibnamefont
  {Yao}}, \bibinfo {author} {\bibfnamefont {D.}~\bibnamefont {Xiao}}, \bibinfo
  {author} {\bibfnamefont {P.}~\bibnamefont {Jarillo-Herrero}},\ and\ \bibinfo
  {author} {\bibfnamefont {X.}~\bibnamefont {Xu}},\ }\bibfield  {title}
  {\bibinfo {title} {Layer-dependent ferromagnetism in a van der {W}aals
  crystal down to the monolayer limit},\ }\href
  {https://doi.org/10.1038/nature22391} {\bibfield  {journal} {\bibinfo
  {journal} {Nature}\ }\textbf {\bibinfo {volume} {546}},\ \bibinfo {pages}
  {270} (\bibinfo {year} {2017})}\BibitemShut {NoStop}%
\bibitem [{\citenamefont {Freitas}\ \emph {et~al.}(2015)\citenamefont
  {Freitas}, \citenamefont {Weht}, \citenamefont {Sulpice}, \citenamefont
  {Remenyi}, \citenamefont {Strobel}, \citenamefont {Gay}, \citenamefont
  {Marcus},\ and\ \citenamefont {N{\'{u}}{\~{n}}ez-Regueiro}}]{2015jpcm}%
  \BibitemOpen
  \bibfield  {author} {\bibinfo {author} {\bibfnamefont {D.~C.}\ \bibnamefont
  {Freitas}}, \bibinfo {author} {\bibfnamefont {R.}~\bibnamefont {Weht}},
  \bibinfo {author} {\bibfnamefont {A.}~\bibnamefont {Sulpice}}, \bibinfo
  {author} {\bibfnamefont {G.}~\bibnamefont {Remenyi}}, \bibinfo {author}
  {\bibfnamefont {P.}~\bibnamefont {Strobel}}, \bibinfo {author} {\bibfnamefont
  {F.}~\bibnamefont {Gay}}, \bibinfo {author} {\bibfnamefont {J.}~\bibnamefont
  {Marcus}},\ and\ \bibinfo {author} {\bibfnamefont {M.}~\bibnamefont
  {N{\'{u}}{\~{n}}ez-Regueiro}},\ }\bibfield  {title} {\bibinfo {title}
  {Ferromagnetism in layered metastable 1${T}$-{CrTe}$_{2}$},\ }\href
  {https://doi.org/10.1088/0953-8984/27/17/176002} {\bibfield  {journal}
  {\bibinfo  {journal} {Journal of Physics: Condensed Matter}\ }\textbf
  {\bibinfo {volume} {27}},\ \bibinfo {pages} {176002} (\bibinfo {year}
  {2015})}\BibitemShut {NoStop}%
\bibitem [{\citenamefont {Sun}\ \emph {et~al.}(2020)\citenamefont {Sun},
  \citenamefont {Li}, \citenamefont {Wang}, \citenamefont {Sui}, \citenamefont
  {Zhang}, \citenamefont {Wang}, \citenamefont {Liu}, \citenamefont {Li},
  \citenamefont {Feng} \emph {et~al.}}]{2020sun}%
  \BibitemOpen
  \bibfield  {author} {\bibinfo {author} {\bibfnamefont {X.}~\bibnamefont
  {Sun}}, \bibinfo {author} {\bibfnamefont {W.}~\bibnamefont {Li}}, \bibinfo
  {author} {\bibfnamefont {X.}~\bibnamefont {Wang}}, \bibinfo {author}
  {\bibfnamefont {Q.}~\bibnamefont {Sui}}, \bibinfo {author} {\bibfnamefont
  {T.}~\bibnamefont {Zhang}}, \bibinfo {author} {\bibfnamefont
  {Z.}~\bibnamefont {Wang}}, \bibinfo {author} {\bibfnamefont {L.}~\bibnamefont
  {Liu}}, \bibinfo {author} {\bibfnamefont {D.}~\bibnamefont {Li}}, \bibinfo
  {author} {\bibfnamefont {S.}~\bibnamefont {Feng}}, \emph {et~al.},\
  }\bibfield  {title} {\bibinfo {title} {Room temperature ferromagnetism in
  ultra-thin van der {W}aals crystals of 1{T}-{CrTe}$_{2}$},\ }\href
  {https://doi.org/10.1007/s12274-020-3021-4} {\bibfield  {journal} {\bibinfo
  {journal} {Nano Research}\ }\textbf {\bibinfo {volume} {13}},\ \bibinfo
  {pages} {3358} (\bibinfo {year} {2020})}\BibitemShut {NoStop}%
\bibitem [{\citenamefont {Deng}\ \emph {et~al.}(2018)\citenamefont {Deng},
  \citenamefont {Yu}, \citenamefont {Song}, \citenamefont {Zhang},
  \citenamefont {Wang}, \citenamefont {Sun}, \citenamefont {Yi}, \citenamefont
  {Wu}, \citenamefont {Wu}, \citenamefont {Zhu} \emph {et~al.}}]{2018deng}%
  \BibitemOpen
  \bibfield  {author} {\bibinfo {author} {\bibfnamefont {Y.}~\bibnamefont
  {Deng}}, \bibinfo {author} {\bibfnamefont {Y.}~\bibnamefont {Yu}}, \bibinfo
  {author} {\bibfnamefont {Y.}~\bibnamefont {Song}}, \bibinfo {author}
  {\bibfnamefont {J.}~\bibnamefont {Zhang}}, \bibinfo {author} {\bibfnamefont
  {N.~Z.}\ \bibnamefont {Wang}}, \bibinfo {author} {\bibfnamefont
  {Z.}~\bibnamefont {Sun}}, \bibinfo {author} {\bibfnamefont {Y.}~\bibnamefont
  {Yi}}, \bibinfo {author} {\bibfnamefont {Y.~Z.}\ \bibnamefont {Wu}}, \bibinfo
  {author} {\bibfnamefont {S.}~\bibnamefont {Wu}}, \bibinfo {author}
  {\bibfnamefont {J.}~\bibnamefont {Zhu}}, \emph {et~al.},\ }\bibfield  {title}
  {\bibinfo {title} {Gate-tunable room-temperature ferromagnetism in
  two-dimensional {Fe}$_{3}${GeTe}$_{2}$},\ }\href
  {https://doi.org/10.1038/s41586-018-0626-9} {\bibfield  {journal} {\bibinfo
  {journal} {Nature}\ }\textbf {\bibinfo {volume} {563}},\ \bibinfo {pages}
  {94} (\bibinfo {year} {2018})}\BibitemShut {NoStop}%
\bibitem [{\citenamefont {Seo}\ \emph {et~al.}(2020)\citenamefont {Seo},
  \citenamefont {Kim}, \citenamefont {An}, \citenamefont {Kim}, \citenamefont
  {Kim}, \citenamefont {Hwang}, \citenamefont {Kim} \emph
  {et~al.}}]{2020sciadv}%
  \BibitemOpen
  \bibfield  {author} {\bibinfo {author} {\bibfnamefont {J.}~\bibnamefont
  {Seo}}, \bibinfo {author} {\bibfnamefont {D.~Y.}\ \bibnamefont {Kim}},
  \bibinfo {author} {\bibfnamefont {E.~S.}\ \bibnamefont {An}}, \bibinfo
  {author} {\bibfnamefont {K.}~\bibnamefont {Kim}}, \bibinfo {author}
  {\bibfnamefont {G.-Y.}\ \bibnamefont {Kim}}, \bibinfo {author} {\bibfnamefont
  {S.-Y.}\ \bibnamefont {Hwang}}, \bibinfo {author} {\bibfnamefont {D.~W.}\
  \bibnamefont {Kim}}, \emph {et~al.},\ }\bibfield  {title} {\bibinfo {title}
  {Nearly room temperature ferromagnetism in a magnetic metal-rich van der
  {W}aals metal},\ }\href {https://doi.org/10.1126/sciadv.aay8912} {\bibfield
  {journal} {\bibinfo  {journal} {Science Advances}\ }\textbf {\bibinfo
  {volume} {6}},\ \bibinfo {pages} {eaay8912} (\bibinfo {year}
  {2020})}\BibitemShut {NoStop}%
\bibitem [{\citenamefont {May}\ \emph {et~al.}(2019)\citenamefont {May},
  \citenamefont {Ovchinnikov}, \citenamefont {Zheng}, \citenamefont {Hermann},
  \citenamefont {Calder}, \citenamefont {Huang}, \citenamefont {Fei},
  \citenamefont {Liu}, \citenamefont {Xu},\ and\ \citenamefont
  {McGuire}}]{2019acsnano}%
  \BibitemOpen
  \bibfield  {author} {\bibinfo {author} {\bibfnamefont {A.~F.}\ \bibnamefont
  {May}}, \bibinfo {author} {\bibfnamefont {D.}~\bibnamefont {Ovchinnikov}},
  \bibinfo {author} {\bibfnamefont {Q.}~\bibnamefont {Zheng}}, \bibinfo
  {author} {\bibfnamefont {R.}~\bibnamefont {Hermann}}, \bibinfo {author}
  {\bibfnamefont {S.}~\bibnamefont {Calder}}, \bibinfo {author} {\bibfnamefont
  {B.}~\bibnamefont {Huang}}, \bibinfo {author} {\bibfnamefont
  {Z.}~\bibnamefont {Fei}}, \bibinfo {author} {\bibfnamefont {Y.}~\bibnamefont
  {Liu}}, \bibinfo {author} {\bibfnamefont {X.}~\bibnamefont {Xu}},\ and\
  \bibinfo {author} {\bibfnamefont {M.~A.}\ \bibnamefont {McGuire}},\
  }\bibfield  {title} {\bibinfo {title} {Ferromagnetism {N}ear {R}oom
  {T}emperature in the {C}leavable van der {W}aals {C}rystal
  {Fe}$_{5}${GeTe}$_{2}$},\ }\href {https://doi.org/10.1021/acsnano.8b09660}
  {\bibfield  {journal} {\bibinfo  {journal} {ACS Nano}\ }\textbf {\bibinfo
  {volume} {13}},\ \bibinfo {pages} {4436} (\bibinfo {year}
  {2019})}\BibitemShut {NoStop}%
\bibitem [{\citenamefont {Stahl}\ \emph {et~al.}(2018)\citenamefont {Stahl},
  \citenamefont {Shlaen},\ and\ \citenamefont {Johrendt}}]{2018za}%
  \BibitemOpen
  \bibfield  {author} {\bibinfo {author} {\bibfnamefont {J.}~\bibnamefont
  {Stahl}}, \bibinfo {author} {\bibfnamefont {E.}~\bibnamefont {Shlaen}},\ and\
  \bibinfo {author} {\bibfnamefont {D.}~\bibnamefont {Johrendt}},\ }\bibfield
  {title} {\bibinfo {title} {The van der waals ferromagnets
  {Fe}$_{5-\delta}${GeTe}$_{2}$ and
  {Fe}$_{5-x-\delta}${Ni}$_{x}${GeTe}$_{2}$–crystal structure, stacking
  faults, and magnetic properties},\ }\href
  {https://doi.org/https://doi.org/10.1002/zaac.201800456} {\bibfield
  {journal} {\bibinfo  {journal} {Z. Anorg. Allg. Chem}\ }\textbf {\bibinfo
  {volume} {644}},\ \bibinfo {pages} {1923} (\bibinfo {year}
  {2018})}\BibitemShut {NoStop}%
\bibitem [{\citenamefont {Zhang}\ \emph {et~al.}(2020)\citenamefont {Zhang},
  \citenamefont {Chen}, \citenamefont {Zhai}, \citenamefont {Chen},
  \citenamefont {Caretta}, \citenamefont {Huang}, \citenamefont {Chopdekar},
  \citenamefont {Cao}, \citenamefont {Sun}, \citenamefont {Yao}, \citenamefont
  {Birgeneau},\ and\ \citenamefont {Ramesh}}]{2020prbzhang}%
  \BibitemOpen
  \bibfield  {author} {\bibinfo {author} {\bibfnamefont {H.}~\bibnamefont
  {Zhang}}, \bibinfo {author} {\bibfnamefont {R.}~\bibnamefont {Chen}},
  \bibinfo {author} {\bibfnamefont {K.}~\bibnamefont {Zhai}}, \bibinfo {author}
  {\bibfnamefont {X.}~\bibnamefont {Chen}}, \bibinfo {author} {\bibfnamefont
  {L.}~\bibnamefont {Caretta}}, \bibinfo {author} {\bibfnamefont
  {X.}~\bibnamefont {Huang}}, \bibinfo {author} {\bibfnamefont {R.~V.}\
  \bibnamefont {Chopdekar}}, \bibinfo {author} {\bibfnamefont {J.}~\bibnamefont
  {Cao}}, \bibinfo {author} {\bibfnamefont {J.}~\bibnamefont {Sun}}, \bibinfo
  {author} {\bibfnamefont {J.}~\bibnamefont {Yao}}, \bibinfo {author}
  {\bibfnamefont {R.}~\bibnamefont {Birgeneau}},\ and\ \bibinfo {author}
  {\bibfnamefont {R.}~\bibnamefont {Ramesh}},\ }\bibfield  {title} {\bibinfo
  {title} {Itinerant ferromagnetism in van der waals {Fe}$_{5-x}${GeTe}$_{2}$
  crystals above room temperature},\ }\href
  {https://doi.org/10.1103/PhysRevB.102.064417} {\bibfield  {journal} {\bibinfo
   {journal} {Phys. Rev. B}\ }\textbf {\bibinfo {volume} {102}},\ \bibinfo
  {pages} {064417} (\bibinfo {year} {2020})}\BibitemShut {NoStop}%
\bibitem [{\citenamefont {Alahmed}\ \emph {et~al.}(2021)\citenamefont
  {Alahmed}, \citenamefont {Nepal}, \citenamefont {Macy}, \citenamefont
  {Zheng}, \citenamefont {Casas}, \citenamefont {Sapkota} \emph
  {et~al.}}]{20212dm}%
  \BibitemOpen
  \bibfield  {author} {\bibinfo {author} {\bibfnamefont {L.}~\bibnamefont
  {Alahmed}}, \bibinfo {author} {\bibfnamefont {B.}~\bibnamefont {Nepal}},
  \bibinfo {author} {\bibfnamefont {J.}~\bibnamefont {Macy}}, \bibinfo {author}
  {\bibfnamefont {W.}~\bibnamefont {Zheng}}, \bibinfo {author} {\bibfnamefont
  {B.}~\bibnamefont {Casas}}, \bibinfo {author} {\bibfnamefont
  {A.}~\bibnamefont {Sapkota}}, \emph {et~al.},\ }\bibfield  {title} {\bibinfo
  {title} {Magnetism and spin dynamics in room-temperature van der waals magnet
  {Fe}$_{5-x}${GeTe}$_{2}$},\ }\href {https://doi.org/10.1088/2053-1583/ac2028}
  {\bibfield  {journal} {\bibinfo  {journal} {2D Materials}\ }\textbf {\bibinfo
  {volume} {8}},\ \bibinfo {pages} {045030} (\bibinfo {year}
  {2021})}\BibitemShut {NoStop}%
\bibitem [{\citenamefont {Chen}\ \emph
  {et~al.}(2022{\natexlab{a}})\citenamefont {Chen}, \citenamefont {Asif},
  \citenamefont {Whalen}, \citenamefont {T{\'{a}}mara-Isaza}, \citenamefont
  {Luetke}, \citenamefont {Wang}, \citenamefont {Wang}, \citenamefont {Ayako},
  \citenamefont {Lamsal}, \citenamefont {May}, \citenamefont {McGuire},
  \citenamefont {Chakraborty}, \citenamefont {Xiao},\ and\ \citenamefont
  {Ku}}]{chen2022}%
  \BibitemOpen
  \bibfield  {author} {\bibinfo {author} {\bibfnamefont {H.}~\bibnamefont
  {Chen}}, \bibinfo {author} {\bibfnamefont {S.}~\bibnamefont {Asif}}, \bibinfo
  {author} {\bibfnamefont {M.}~\bibnamefont {Whalen}}, \bibinfo {author}
  {\bibfnamefont {J.}~\bibnamefont {T{\'{a}}mara-Isaza}}, \bibinfo {author}
  {\bibfnamefont {B.}~\bibnamefont {Luetke}}, \bibinfo {author} {\bibfnamefont
  {Y.}~\bibnamefont {Wang}}, \bibinfo {author} {\bibfnamefont {X.}~\bibnamefont
  {Wang}}, \bibinfo {author} {\bibfnamefont {M.}~\bibnamefont {Ayako}},
  \bibinfo {author} {\bibfnamefont {S.}~\bibnamefont {Lamsal}}, \bibinfo
  {author} {\bibfnamefont {A.~F.}\ \bibnamefont {May}}, \bibinfo {author}
  {\bibfnamefont {M.~A.}\ \bibnamefont {McGuire}}, \bibinfo {author}
  {\bibfnamefont {C.}~\bibnamefont {Chakraborty}}, \bibinfo {author}
  {\bibfnamefont {J.~Q.}\ \bibnamefont {Xiao}},\ and\ \bibinfo {author}
  {\bibfnamefont {M.~J.~H.}\ \bibnamefont {Ku}},\ }\bibfield  {title} {\bibinfo
  {title} {Revealing room temperature ferromagnetism in exfoliated
  {F}e$_5${G}e{T}e$_2$ flakes with quantum magnetic imaging},\ }\href
  {https://doi.org/10.1088/2053-1583/ac57a9} {\bibfield  {journal} {\bibinfo
  {journal} {2D Materials}\ }\textbf {\bibinfo {volume} {9}},\ \bibinfo {pages}
  {025017} (\bibinfo {year} {2022}{\natexlab{a}})}\BibitemShut {NoStop}%
\bibitem [{\citenamefont {May}\ \emph {et~al.}(2021)\citenamefont {May},
  \citenamefont {Yan}, \citenamefont {Hermann}, \citenamefont {Du},\ and\
  \citenamefont {McGuire}}]{2021may}%
  \BibitemOpen
  \bibfield  {author} {\bibinfo {author} {\bibfnamefont {A.~F.}\ \bibnamefont
  {May}}, \bibinfo {author} {\bibfnamefont {J.}~\bibnamefont {Yan}}, \bibinfo
  {author} {\bibfnamefont {R.}~\bibnamefont {Hermann}}, \bibinfo {author}
  {\bibfnamefont {M.-H.}\ \bibnamefont {Du}},\ and\ \bibinfo {author}
  {\bibfnamefont {M.~A.}\ \bibnamefont {McGuire}},\ }\bibfield  {title}
  {\bibinfo {title} {Tuning the room temperature ferromagnetism in
  {Fe}$_{5}${GeTe}$_{2}$ by arsenic substitution},\ }\href
  {https://doi.org/10.1088/2053-1583/ac34d9} {\bibfield  {journal} {\bibinfo
  {journal} {2D Materials}\ }\textbf {\bibinfo {volume} {9}},\ \bibinfo {pages}
  {015013} (\bibinfo {year} {2021})}\BibitemShut {NoStop}%
\bibitem [{\citenamefont {Chen}\ \emph
  {et~al.}(2022{\natexlab{b}})\citenamefont {Chen}, \citenamefont {Shao},
  \citenamefont {Chen}, \citenamefont {Susarla}, \citenamefont {Hogan},
  \citenamefont {He}, \citenamefont {Zhang}, \citenamefont {Wang},
  \citenamefont {Yao}, \citenamefont {Ercius}, \citenamefont {Muller},
  \citenamefont {Ramesh},\ and\ \citenamefont {Birgeneau}}]{2022prl}%
  \BibitemOpen
  \bibfield  {author} {\bibinfo {author} {\bibfnamefont {X.}~\bibnamefont
  {Chen}}, \bibinfo {author} {\bibfnamefont {Y.-T.}\ \bibnamefont {Shao}},
  \bibinfo {author} {\bibfnamefont {R.}~\bibnamefont {Chen}}, \bibinfo {author}
  {\bibfnamefont {S.}~\bibnamefont {Susarla}}, \bibinfo {author} {\bibfnamefont
  {T.}~\bibnamefont {Hogan}}, \bibinfo {author} {\bibfnamefont
  {Y.}~\bibnamefont {He}}, \bibinfo {author} {\bibfnamefont {H.}~\bibnamefont
  {Zhang}}, \bibinfo {author} {\bibfnamefont {S.}~\bibnamefont {Wang}},
  \bibinfo {author} {\bibfnamefont {J.}~\bibnamefont {Yao}}, \bibinfo {author}
  {\bibfnamefont {P.}~\bibnamefont {Ercius}}, \bibinfo {author} {\bibfnamefont
  {D.~A.}\ \bibnamefont {Muller}}, \bibinfo {author} {\bibfnamefont
  {R.}~\bibnamefont {Ramesh}},\ and\ \bibinfo {author} {\bibfnamefont {R.~J.}\
  \bibnamefont {Birgeneau}},\ }\bibfield  {title} {\bibinfo {title} {Pervasive
  beyond room-temperature ferromagnetism in a doped van der waals magnet},\
  }\href {https://doi.org/10.1103/PhysRevLett.128.217203} {\bibfield  {journal}
  {\bibinfo  {journal} {Phys. Rev. Lett.}\ }\textbf {\bibinfo {volume} {128}},\
  \bibinfo {pages} {217203} (\bibinfo {year} {2022}{\natexlab{b}})}\BibitemShut
  {NoStop}%
\bibitem [{\citenamefont {Zhang}\ \emph
  {et~al.}(2022{\natexlab{a}})\citenamefont {Zhang}, \citenamefont {Raftrey},
  \citenamefont {Chan}, \citenamefont {Shao}, \citenamefont {Chen},
  \citenamefont {Chen}, \citenamefont {Huang}, \citenamefont {Reichanadter},
  \citenamefont {Dong}, \citenamefont {Susarla}, \citenamefont {Caretta},
  \citenamefont {Chen}, \citenamefont {Yao}, \citenamefont {Fischer},
  \citenamefont {Neaton}, \citenamefont {Wu}, \citenamefont {Muller},
  \citenamefont {Birgeneau},\ and\ \citenamefont {Ramesh}}]{2022sciav}%
  \BibitemOpen
  \bibfield  {author} {\bibinfo {author} {\bibfnamefont {H.}~\bibnamefont
  {Zhang}}, \bibinfo {author} {\bibfnamefont {D.}~\bibnamefont {Raftrey}},
  \bibinfo {author} {\bibfnamefont {Y.-T.}\ \bibnamefont {Chan}}, \bibinfo
  {author} {\bibfnamefont {Y.-T.}\ \bibnamefont {Shao}}, \bibinfo {author}
  {\bibfnamefont {R.}~\bibnamefont {Chen}}, \bibinfo {author} {\bibfnamefont
  {X.}~\bibnamefont {Chen}}, \bibinfo {author} {\bibfnamefont {X.}~\bibnamefont
  {Huang}}, \bibinfo {author} {\bibfnamefont {J.~T.}\ \bibnamefont
  {Reichanadter}}, \bibinfo {author} {\bibfnamefont {K.}~\bibnamefont {Dong}},
  \bibinfo {author} {\bibfnamefont {S.}~\bibnamefont {Susarla}}, \bibinfo
  {author} {\bibfnamefont {L.}~\bibnamefont {Caretta}}, \bibinfo {author}
  {\bibfnamefont {Z.}~\bibnamefont {Chen}}, \bibinfo {author} {\bibfnamefont
  {J.}~\bibnamefont {Yao}}, \bibinfo {author} {\bibfnamefont {P.}~\bibnamefont
  {Fischer}}, \bibinfo {author} {\bibfnamefont {J.~B.}\ \bibnamefont {Neaton}},
  \bibinfo {author} {\bibfnamefont {W.}~\bibnamefont {Wu}}, \bibinfo {author}
  {\bibfnamefont {D.~A.}\ \bibnamefont {Muller}}, \bibinfo {author}
  {\bibfnamefont {R.~J.}\ \bibnamefont {Birgeneau}},\ and\ \bibinfo {author}
  {\bibfnamefont {R.}~\bibnamefont {Ramesh}},\ }\bibfield  {title} {\bibinfo
  {title} {Room-temperature skyrmion lattice in a layered magnet
  {Fe}$_{0.5}${Co}$_{0.5}${GeTe}$_{2}$},\ }\href
  {https://doi.org/10.1126/sciadv.abm7103} {\bibfield  {journal} {\bibinfo
  {journal} {Science Advances}\ }\textbf {\bibinfo {volume} {8}},\ \bibinfo
  {pages} {eabm7103} (\bibinfo {year} {2022}{\natexlab{a}})}\BibitemShut
  {NoStop}%
\bibitem [{\citenamefont {Zhang}\ \emph
  {et~al.}(2022{\natexlab{b}})\citenamefont {Zhang}, \citenamefont {Shao},
  \citenamefont {Chen}, \citenamefont {Chen}, \citenamefont {Susarla},
  \citenamefont {Raftrey}, \citenamefont {Reichanadter}, \citenamefont
  {Caretta}, \citenamefont {Huang}, \citenamefont {Settineri}, \citenamefont
  {Chen}, \citenamefont {Zhou}, \citenamefont {Bourret-Courchesne},
  \citenamefont {Ercius}, \citenamefont {Yao}, \citenamefont {Fischer},
  \citenamefont {Neaton}, \citenamefont {Muller}, \citenamefont {Birgeneau},\
  and\ \citenamefont {Ramesh}}]{2022prmz}%
  \BibitemOpen
  \bibfield  {author} {\bibinfo {author} {\bibfnamefont {H.}~\bibnamefont
  {Zhang}}, \bibinfo {author} {\bibfnamefont {Y.-T.}\ \bibnamefont {Shao}},
  \bibinfo {author} {\bibfnamefont {R.}~\bibnamefont {Chen}}, \bibinfo {author}
  {\bibfnamefont {X.}~\bibnamefont {Chen}}, \bibinfo {author} {\bibfnamefont
  {S.}~\bibnamefont {Susarla}}, \bibinfo {author} {\bibfnamefont
  {D.}~\bibnamefont {Raftrey}}, \bibinfo {author} {\bibfnamefont {J.~T.}\
  \bibnamefont {Reichanadter}}, \bibinfo {author} {\bibfnamefont
  {L.}~\bibnamefont {Caretta}}, \bibinfo {author} {\bibfnamefont
  {X.}~\bibnamefont {Huang}}, \bibinfo {author} {\bibfnamefont {N.~S.}\
  \bibnamefont {Settineri}}, \bibinfo {author} {\bibfnamefont {Z.}~\bibnamefont
  {Chen}}, \bibinfo {author} {\bibfnamefont {J.}~\bibnamefont {Zhou}}, \bibinfo
  {author} {\bibfnamefont {E.}~\bibnamefont {Bourret-Courchesne}}, \bibinfo
  {author} {\bibfnamefont {P.}~\bibnamefont {Ercius}}, \bibinfo {author}
  {\bibfnamefont {J.}~\bibnamefont {Yao}}, \bibinfo {author} {\bibfnamefont
  {P.}~\bibnamefont {Fischer}}, \bibinfo {author} {\bibfnamefont {J.~B.}\
  \bibnamefont {Neaton}}, \bibinfo {author} {\bibfnamefont {D.~A.}\
  \bibnamefont {Muller}}, \bibinfo {author} {\bibfnamefont {R.~J.}\
  \bibnamefont {Birgeneau}},\ and\ \bibinfo {author} {\bibfnamefont
  {R.}~\bibnamefont {Ramesh}},\ }\bibfield  {title} {\bibinfo {title} {A room
  temperature polar magnetic metal},\ }\href
  {https://doi.org/10.1103/PhysRevMaterials.6.044403} {\bibfield  {journal}
  {\bibinfo  {journal} {Phys. Rev. Materials}\ }\textbf {\bibinfo {volume}
  {6}},\ \bibinfo {pages} {044403} (\bibinfo {year}
  {2022}{\natexlab{b}})}\BibitemShut {NoStop}%
\bibitem [{\citenamefont {Joe}\ \emph {et~al.}(2019)\citenamefont {Joe},
  \citenamefont {Yang},\ and\ \citenamefont {Lee}}]{2019first}%
  \BibitemOpen
  \bibfield  {author} {\bibinfo {author} {\bibfnamefont {M.}~\bibnamefont
  {Joe}}, \bibinfo {author} {\bibfnamefont {U.}~\bibnamefont {Yang}},\ and\
  \bibinfo {author} {\bibfnamefont {C.}~\bibnamefont {Lee}},\ }\bibfield
  {title} {\bibinfo {title} {First-principles study of ferromagnetic metal
  {F}e$_{5}${GeTe}$_{2}$},\ }\href
  {https://doi.org/https://doi.org/10.1016/j.nanoms.2019.09.009} {\bibfield
  {journal} {\bibinfo  {journal} {Nano Materials Science}\ }\textbf {\bibinfo
  {volume} {1}},\ \bibinfo {pages} {299} (\bibinfo {year} {2019})}\BibitemShut
  {NoStop}%
\bibitem [{\citenamefont {Tan}\ \emph {et~al.}(2021)\citenamefont {Tan},
  \citenamefont {Xie}, \citenamefont {Zheng}, \citenamefont {Aloufi},
  \citenamefont {Albarakati}, \citenamefont {Algarni}, \citenamefont {Li},
  \citenamefont {Partridge}, \citenamefont {Culcer}, \citenamefont {Wang},
  \citenamefont {Yi}, \citenamefont {Tian}, \citenamefont {Xiong},
  \citenamefont {Zhao},\ and\ \citenamefont {Wang}}]{2021zhaoyu}%
  \BibitemOpen
  \bibfield  {author} {\bibinfo {author} {\bibfnamefont {C.}~\bibnamefont
  {Tan}}, \bibinfo {author} {\bibfnamefont {W.-Q.}\ \bibnamefont {Xie}},
  \bibinfo {author} {\bibfnamefont {G.}~\bibnamefont {Zheng}}, \bibinfo
  {author} {\bibfnamefont {N.}~\bibnamefont {Aloufi}}, \bibinfo {author}
  {\bibfnamefont {S.}~\bibnamefont {Albarakati}}, \bibinfo {author}
  {\bibfnamefont {M.}~\bibnamefont {Algarni}}, \bibinfo {author} {\bibfnamefont
  {J.}~\bibnamefont {Li}}, \bibinfo {author} {\bibfnamefont {J.}~\bibnamefont
  {Partridge}}, \bibinfo {author} {\bibfnamefont {D.}~\bibnamefont {Culcer}},
  \bibinfo {author} {\bibfnamefont {X.}~\bibnamefont {Wang}}, \bibinfo {author}
  {\bibfnamefont {J.~B.}\ \bibnamefont {Yi}}, \bibinfo {author} {\bibfnamefont
  {M.}~\bibnamefont {Tian}}, \bibinfo {author} {\bibfnamefont {Y.}~\bibnamefont
  {Xiong}}, \bibinfo {author} {\bibfnamefont {Y.-J.}\ \bibnamefont {Zhao}},\
  and\ \bibinfo {author} {\bibfnamefont {L.}~\bibnamefont {Wang}},\ }\bibfield
  {title} {\bibinfo {title} {Gate-controlled magnetic phase transition in a van
  der waals magnet {F}e5{G}e{T}e$_{2}$},\ }\href
  {https://doi.org/10.1021/acs.nanolett.1c01108} {\bibfield  {journal}
  {\bibinfo  {journal} {Nano Letters}\ }\textbf {\bibinfo {volume} {21}},\
  \bibinfo {pages} {5599} (\bibinfo {year} {2021})}\BibitemShut {NoStop}%
\bibitem [{\citenamefont {Liu}\ \emph {et~al.}(2022)\citenamefont {Liu},
  \citenamefont {Xing}, \citenamefont {Jiang}, \citenamefont {Guo},
  \citenamefont {Jiang}, \citenamefont {Qi},\ and\ \citenamefont
  {Zhao}}]{2022liu}%
  \BibitemOpen
  \bibfield  {author} {\bibinfo {author} {\bibfnamefont {Q.}~\bibnamefont
  {Liu}}, \bibinfo {author} {\bibfnamefont {J.}~\bibnamefont {Xing}}, \bibinfo
  {author} {\bibfnamefont {Z.}~\bibnamefont {Jiang}}, \bibinfo {author}
  {\bibfnamefont {Y.}~\bibnamefont {Guo}}, \bibinfo {author} {\bibfnamefont
  {X.}~\bibnamefont {Jiang}}, \bibinfo {author} {\bibfnamefont
  {Y.}~\bibnamefont {Qi}},\ and\ \bibinfo {author} {\bibfnamefont
  {J.}~\bibnamefont {Zhao}},\ }\bibfield  {title} {\bibinfo {title}
  {Layer-dependent magnetic phase diagram in {F}e$_{n}${GeTe}$_{2}$ (3 $\leq $
  n $\leq $ 7) ultrathin films},\ }\href
  {https://doi.org/10.1126/sciadv.abm7103} {\bibfield  {journal} {\bibinfo
  {journal} {Communications Physics}\ }\textbf {\bibinfo {volume} {5}},\
  \bibinfo {pages} {1} (\bibinfo {year} {2022})}\BibitemShut {NoStop}%
\bibitem [{\citenamefont {Yang}\ \emph {et~al.}(2021)\citenamefont {Yang},
  \citenamefont {Zhou}, \citenamefont {Feng},\ and\ \citenamefont
  {Yao}}]{2021prbyao}%
  \BibitemOpen
  \bibfield  {author} {\bibinfo {author} {\bibfnamefont {X.}~\bibnamefont
  {Yang}}, \bibinfo {author} {\bibfnamefont {X.}~\bibnamefont {Zhou}}, \bibinfo
  {author} {\bibfnamefont {W.}~\bibnamefont {Feng}},\ and\ \bibinfo {author}
  {\bibfnamefont {Y.}~\bibnamefont {Yao}},\ }\bibfield  {title} {\bibinfo
  {title} {Strong magneto-optical effect and anomalous transport in the
  two-dimensional van der waals magnets {Fe}$_{n}${GeTe}$_{2}$ ($n=3$, 4, 5)},\
  }\href {https://doi.org/10.1103/PhysRevB.104.104427} {\bibfield  {journal}
  {\bibinfo  {journal} {Phys. Rev. B}\ }\textbf {\bibinfo {volume} {104}},\
  \bibinfo {pages} {104427} (\bibinfo {year} {2021})}\BibitemShut {NoStop}%
\bibitem [{\citenamefont {Ly}\ \emph {et~al.}(2021)\citenamefont {Ly},
  \citenamefont {Park}, \citenamefont {Kim}, \citenamefont {Ahn}, \citenamefont
  {Lee}, \citenamefont {Kim}, \citenamefont {Park}, \citenamefont {Duvjir},
  \citenamefont {Lam}, \citenamefont {Jang}, \citenamefont {You}, \citenamefont
  {Jo}, \citenamefont {Kim}, \citenamefont {Lee}, \citenamefont {Kim},\ and\
  \citenamefont {Kim}}]{2021adfm}%
  \BibitemOpen
  \bibfield  {author} {\bibinfo {author} {\bibfnamefont {T.~T.}\ \bibnamefont
  {Ly}}, \bibinfo {author} {\bibfnamefont {J.}~\bibnamefont {Park}}, \bibinfo
  {author} {\bibfnamefont {K.}~\bibnamefont {Kim}}, \bibinfo {author}
  {\bibfnamefont {H.-B.}\ \bibnamefont {Ahn}}, \bibinfo {author} {\bibfnamefont
  {N.~J.}\ \bibnamefont {Lee}}, \bibinfo {author} {\bibfnamefont
  {K.}~\bibnamefont {Kim}}, \bibinfo {author} {\bibfnamefont {T.-E.}\
  \bibnamefont {Park}}, \bibinfo {author} {\bibfnamefont {G.}~\bibnamefont
  {Duvjir}}, \bibinfo {author} {\bibfnamefont {N.~H.}\ \bibnamefont {Lam}},
  \bibinfo {author} {\bibfnamefont {K.}~\bibnamefont {Jang}}, \bibinfo {author}
  {\bibfnamefont {C.-Y.}\ \bibnamefont {You}}, \bibinfo {author} {\bibfnamefont
  {Y.}~\bibnamefont {Jo}}, \bibinfo {author} {\bibfnamefont {S.~K.}\
  \bibnamefont {Kim}}, \bibinfo {author} {\bibfnamefont {C.}~\bibnamefont
  {Lee}}, \bibinfo {author} {\bibfnamefont {S.}~\bibnamefont {Kim}},\ and\
  \bibinfo {author} {\bibfnamefont {J.}~\bibnamefont {Kim}},\ }\bibfield
  {title} {\bibinfo {title} {Direct observation of {Fe}-{Ge} ordering in
  {F}e$_{5-x}${GeTe}$_{2}$ crystals and resultant helimagnetism},\ }\href
  {https://doi.org/https://doi.org/10.1002/adfm.202009758} {\bibfield
  {journal} {\bibinfo  {journal} {Advanced Functional Materials}\ }\textbf
  {\bibinfo {volume} {31}},\ \bibinfo {pages} {2009758} (\bibinfo {year}
  {2021})}\BibitemShut {NoStop}%
\bibitem [{\citenamefont {Wu}\ \emph {et~al.}(2021)\citenamefont {Wu},
  \citenamefont {Lei}, \citenamefont {Yin}, \citenamefont {Zhao}, \citenamefont
  {Li}, \citenamefont {Wang}, \citenamefont {Liu}, \citenamefont {Song},
  \citenamefont {Ma}, \citenamefont {Ding}, \citenamefont {Cheng},
  \citenamefont {Liu}, \citenamefont {Lei},\ and\ \citenamefont
  {Wang}}]{2021prb}%
  \BibitemOpen
  \bibfield  {author} {\bibinfo {author} {\bibfnamefont {X.}~\bibnamefont
  {Wu}}, \bibinfo {author} {\bibfnamefont {L.}~\bibnamefont {Lei}}, \bibinfo
  {author} {\bibfnamefont {Q.}~\bibnamefont {Yin}}, \bibinfo {author}
  {\bibfnamefont {N.-N.}\ \bibnamefont {Zhao}}, \bibinfo {author}
  {\bibfnamefont {M.}~\bibnamefont {Li}}, \bibinfo {author} {\bibfnamefont
  {Z.}~\bibnamefont {Wang}}, \bibinfo {author} {\bibfnamefont {Q.}~\bibnamefont
  {Liu}}, \bibinfo {author} {\bibfnamefont {W.}~\bibnamefont {Song}}, \bibinfo
  {author} {\bibfnamefont {H.}~\bibnamefont {Ma}}, \bibinfo {author}
  {\bibfnamefont {P.}~\bibnamefont {Ding}}, \bibinfo {author} {\bibfnamefont
  {Z.}~\bibnamefont {Cheng}}, \bibinfo {author} {\bibfnamefont
  {K.}~\bibnamefont {Liu}}, \bibinfo {author} {\bibfnamefont {H.}~\bibnamefont
  {Lei}},\ and\ \bibinfo {author} {\bibfnamefont {S.}~\bibnamefont {Wang}},\
  }\bibfield  {title} {\bibinfo {title} {Direct observation of competition
  between charge order and itinerant ferromagnetism in the van der waals
  crystal {Fe}$_{5-x}${GeTe}$_{2}$},\ }\href
  {https://doi.org/10.1103/PhysRevB.104.165101} {\bibfield  {journal} {\bibinfo
   {journal} {Phys. Rev. B}\ }\textbf {\bibinfo {volume} {104}},\ \bibinfo
  {pages} {165101} (\bibinfo {year} {2021})}\BibitemShut {NoStop}%
\bibitem [{\citenamefont {Ershadrad}\ \emph {et~al.}(2022)\citenamefont
  {Ershadrad}, \citenamefont {Ghosh}, \citenamefont {Wang}, \citenamefont
  {Kvashnin},\ and\ \citenamefont {Sanyal}}]{2022jpcl}%
  \BibitemOpen
  \bibfield  {author} {\bibinfo {author} {\bibfnamefont {S.}~\bibnamefont
  {Ershadrad}}, \bibinfo {author} {\bibfnamefont {S.}~\bibnamefont {Ghosh}},
  \bibinfo {author} {\bibfnamefont {D.}~\bibnamefont {Wang}}, \bibinfo {author}
  {\bibfnamefont {Y.}~\bibnamefont {Kvashnin}},\ and\ \bibinfo {author}
  {\bibfnamefont {B.}~\bibnamefont {Sanyal}},\ }\bibfield  {title} {\bibinfo
  {title} {Unusual magnetic features in two-dimensional
  {F}e$_{5-x}${GeTe}$_{2}$ induced by structural reconstructions},\ }\href
  {https://doi.org/10.1021/acs.jpclett.2c00692} {\bibfield  {journal} {\bibinfo
   {journal} {The Journal of Physical Chemistry Letters}\ }\textbf {\bibinfo
  {volume} {13}},\ \bibinfo {pages} {4877} (\bibinfo {year}
  {2022})}\BibitemShut {NoStop}%
\bibitem [{\citenamefont {Wannier}(1950)}]{1950prv}%
  \BibitemOpen
  \bibfield  {author} {\bibinfo {author} {\bibfnamefont {G.~H.}\ \bibnamefont
  {Wannier}},\ }\bibfield  {title} {\bibinfo {title} {Antiferromagnetism. {T}he
  {T}riangular {I}sing net},\ }\href {https://doi.org/10.1103/PhysRev.79.357}
  {\bibfield  {journal} {\bibinfo  {journal} {Phys. Rev.}\ }\textbf {\bibinfo
  {volume} {79}},\ \bibinfo {pages} {357} (\bibinfo {year} {1950})}\BibitemShut
  {NoStop}%
\bibitem [{\citenamefont {Saito}\ and\ \citenamefont {Igeta}(1984)}]{1984jpsj}%
  \BibitemOpen
  \bibfield  {author} {\bibinfo {author} {\bibfnamefont {Y.}~\bibnamefont
  {Saito}}\ and\ \bibinfo {author} {\bibfnamefont {K.}~\bibnamefont {Igeta}},\
  }\bibfield  {title} {\bibinfo {title} {Antiferromagnetic {Ising Model on a
  Triangular Lattice}},\ }\href {https://doi.org/10.1143/JPSJ.53.3060}
  {\bibfield  {journal} {\bibinfo  {journal} {Journal of the Physical Society
  of Japan}\ }\textbf {\bibinfo {volume} {53}},\ \bibinfo {pages} {3060}
  (\bibinfo {year} {1984})}\BibitemShut {NoStop}%
\bibitem [{\citenamefont {Ma}\ and\ \citenamefont {Dudarev}(2012)}]{prbma}%
  \BibitemOpen
  \bibfield  {author} {\bibinfo {author} {\bibfnamefont {P.-W.}\ \bibnamefont
  {Ma}}\ and\ \bibinfo {author} {\bibfnamefont {S.~L.}\ \bibnamefont
  {Dudarev}},\ }\bibfield  {title} {\bibinfo {title} {Longitudinal magnetic
  fluctuations in {L}angevin spin dynamics},\ }\href
  {https://doi.org/10.1103/PhysRevB.86.054416} {\bibfield  {journal} {\bibinfo
  {journal} {Phys. Rev. B}\ }\textbf {\bibinfo {volume} {86}},\ \bibinfo
  {pages} {054416} (\bibinfo {year} {2012})}\BibitemShut {NoStop}%
\bibitem [{\citenamefont {Ruban}\ \emph {et~al.}(2007)\citenamefont {Ruban},
  \citenamefont {Khmelevskyi}, \citenamefont {Mohn},\ and\ \citenamefont
  {Johansson}}]{2007prbr}%
  \BibitemOpen
  \bibfield  {author} {\bibinfo {author} {\bibfnamefont {A.~V.}\ \bibnamefont
  {Ruban}}, \bibinfo {author} {\bibfnamefont {S.}~\bibnamefont {Khmelevskyi}},
  \bibinfo {author} {\bibfnamefont {P.}~\bibnamefont {Mohn}},\ and\ \bibinfo
  {author} {\bibfnamefont {B.}~\bibnamefont {Johansson}},\ }\bibfield  {title}
  {\bibinfo {title} {Temperature-induced longitudinal spin fluctuations in {Fe}
  and {Ni}},\ }\href {https://doi.org/10.1103/PhysRevB.75.054402} {\bibfield
  {journal} {\bibinfo  {journal} {Phys. Rev. B}\ }\textbf {\bibinfo {volume}
  {75}},\ \bibinfo {pages} {054402} (\bibinfo {year} {2007})}\BibitemShut
  {NoStop}%
\bibitem [{\citenamefont {Rosengaard}\ and\ \citenamefont
  {Johansson}(1997)}]{97prbr}%
  \BibitemOpen
  \bibfield  {author} {\bibinfo {author} {\bibfnamefont {N.~M.}\ \bibnamefont
  {Rosengaard}}\ and\ \bibinfo {author} {\bibfnamefont {B.}~\bibnamefont
  {Johansson}},\ }\bibfield  {title} {\bibinfo {title} {Finite-temperature
  study of itinerant ferromagnetism in {Fe}, {Co}, and {Ni}},\ }\href
  {https://doi.org/10.1103/PhysRevB.55.14975} {\bibfield  {journal} {\bibinfo
  {journal} {Phys. Rev. B}\ }\textbf {\bibinfo {volume} {55}},\ \bibinfo
  {pages} {14975} (\bibinfo {year} {1997})}\BibitemShut {NoStop}%
\bibitem [{\citenamefont {Uhl}\ and\ \citenamefont {K\"ubler}(1996)}]{97prl}%
  \BibitemOpen
  \bibfield  {author} {\bibinfo {author} {\bibfnamefont {M.}~\bibnamefont
  {Uhl}}\ and\ \bibinfo {author} {\bibfnamefont {J.}~\bibnamefont {K\"ubler}},\
  }\bibfield  {title} {\bibinfo {title} {Exchange-{C}oupled
  {S}pin-{F}luctuation {T}heory: {A}pplication to {F}e, {C}o, and {N}i},\
  }\href {https://doi.org/10.1103/PhysRevLett.77.334} {\bibfield  {journal}
  {\bibinfo  {journal} {Phys. Rev. Lett.}\ }\textbf {\bibinfo {volume} {77}},\
  \bibinfo {pages} {334} (\bibinfo {year} {1996})}\BibitemShut {NoStop}%
\bibitem [{\citenamefont {He}\ \emph {et~al.}(2021)\citenamefont {He},
  \citenamefont {Helbig}, \citenamefont {Verstraete},\ and\ \citenamefont
  {Bousquet}}]{2021cpc}%
  \BibitemOpen
  \bibfield  {author} {\bibinfo {author} {\bibfnamefont {X.}~\bibnamefont
  {He}}, \bibinfo {author} {\bibfnamefont {N.}~\bibnamefont {Helbig}}, \bibinfo
  {author} {\bibfnamefont {M.~J.}\ \bibnamefont {Verstraete}},\ and\ \bibinfo
  {author} {\bibfnamefont {E.}~\bibnamefont {Bousquet}},\ }\bibfield  {title}
  {\bibinfo {title} {{TB2J}: A python package for computing magnetic
  interaction parameters},\ }\href
  {https://doi.org/https://doi.org/10.1016/j.cpc.2021.107938} {\bibfield
  {journal} {\bibinfo  {journal} {Computer Physics Communications}\ }\textbf
  {\bibinfo {volume} {264}},\ \bibinfo {pages} {107938} (\bibinfo {year}
  {2021})}\BibitemShut {NoStop}%
\end{thebibliography}%
\end{document}